\documentclass[%
 aip,
 jmp,
 amsmath,amssymb,
 reprint,%
]{revtex4-1}

\usepackage{graphicx}
\usepackage{dcolumn}
\usepackage{bm}

\usepackage[utf8]{inputenc}
\usepackage[T1]{fontenc}
\usepackage{etoolbox}
\usepackage{mathrsfs}
\usepackage{yfonts}

\usepackage{mathabx}    
\usepackage{xcolor}

\makeatletter
\def\@email#1#2{%
 \endgroup
 \patchcmd{\titleblock@produce}
  {\frontmatter@RRAPformat}
  {\frontmatter@RRAPformat{\produce@RRAP{*#1\href{mailto:#2}{#2}}}\frontmatter@RRAPformat}
  {}{}
}%
\makeatother
\begin{document}

\preprint{AIP/123-QED}

\title{The cosmic multipoles: a series of consistency tests for the Szekeres cosmological model} 



\author{Marie-No\"elle C\'el\'erier}
\affiliation{Laboratoire d'\'etude de l’Univers et des ph\'enom\`enes eXtr\`emes (LUX), Observatoire de Paris-PSL, UMR 8262 CNRS, Sorbonne Universit\'e, 5, Place Jules Janssen, F-92190 Meudon, France}
\email{marie-noelle.celerier@obspm.fr}

\date{\today}

\begin{abstract}
An important property of the quasi-spherical Szekeres cosmological model appears to be widely misunderstood in the literature. In this paper, the claim that the matter density can be fully decomposed on a given 2-shell into a dipole superposed on a monopole is shown to be inaccurate and exact propositions are substituted. Then, the multipole decompositions of three different quantities linked to the mass density are displayed and proposed to constrain the Szekeres parameters with observed data. A test of the axial symmetry assumption, together with that of an on-axis observer, using a property of the angular coefficients is also suggested.
\end{abstract}


\maketitle 

\section{Introduction} \label{intro}

The Cosmic Dipole Anomaly has recently been the subject of a renewed interest in the literature, see \cite{B18,B19,Se21,Si21,D22,S22,D23,W23,T24,S24,M24,W24,vH24,LS25,W25,Sa25,S24c,S25d,B26} and \cite{S25b} for a review. This is due to the display of large radio-galaxy and quasar catalogs, which allow tests inspired by Ellis and Baldwin's proposal \cite{E84} to be performed.

Indeed, these authors noticed that the Cosmological Principle (CP) is not a physical principle, but a mere assumption and, as such, that it needs to be put to the test by observations. They pointed out that, if the CP were to hold, the rest frame of distant sources would coincide with that of the Cosmic Microwave Background (CMB). Moreover, they showed that the comparison of the kinematic dipole issued from our motion with respect to the CMB rest frame with the dipole measured at cosmological scales in the galaxy number counts could be a test of the assumption that the evolution of the Universe might be handled by a time only dependent scale factor $a(t)$. Since such a scale factor is a basic feature of Friedmann-Lema\^itre-Robertson-Walker (FLRW) models, such models could thus be put to the test, and so would be their funding assumption, the CP.

The Ellis-Baldwin test, in its initial form, was only designed to test the CP through $a(t)$ without saying anything else about the underlying cosmology. Indeed, two special relativistic effects were combined for this purpose, the Doppler boosting and the aberration, and the derivation of the kinematic part of the galaxy dipole was done in terms of features measured from a sky projected source sample. Therefore, the test was able to discriminate between FLRW and something else, but the something else could not be determined by the test in its raw version. 

Now, this test has been completed a large number of times, by using different source samples, and the conclusions seem to converge toward the following statement: the direction of the measured matter dipole roughly coincides with that of the CMB dipole, but its amplitude is subsequently larger, with a significance up to $\sim 5 \sigma$. Individual data set significances range from 3.9$\sigma$ (WISE alone \cite{Se21}) to 5.2$\sigma$ (joint NVSS+CatWISE \cite{W24}), with most independent confirmations (RACS \cite{W23}, joint Planck comparisons \cite{W24}) clustering around 4.8–5$\sigma$. The most recent review paper \cite{S25b} claims that the overall tension reachs a $\sim 6.4\sigma$ significance.

All these analyses appear to converge toward the existence of an intrinsic dipole completing the kinematic dipole in the CMB dipole direction. Therefore, the Szekeres solution, which has been recently revived as an appealing model to represent the late universe \cite{C24,C26} and which exhibits a natural intrinsic dipole, might be considered as a possible candidate to solve the Cosmic Dipole Anomaly. 

The Szekeres dipole has been the subject of several studies since the display of its first version by Peter Szekeres himself \cite{S75a}. However, even though this first proposal was perfectly correct, a further misleading claim appeared in the literature \cite{dS85} and was propagated since then under an improper variant \cite{H02,B06,B09,BK10,K16,K18,K20,P24}, of the kind: "In the quasi-spherical Szekeres model the distribution of mass over a surface of constant $t$ and $r$ has the form of a mass-dipole superposed on a monopole". Applied to the mass density $\rho$ of the spacetime stress-energy tensor, this statement is fully improper as we will show in this paper. 

After showing that the quasi-spherical class of the Szekeres (QSS) spacetime is the best one designed for a cosmological use, we will discuss the way different physical densities pertaining to QSS models can be angularly decomposed. Then, we will propose experiments aiming at testing the consistency of the Szekeres models, by using their angular properties.

Since, it has been recently found that the expansion of the Universe exhibits an axial symmetry also compatible with the direction of the CMB dipole \cite{B16,C19,R25,K26}, we will propose to test systematically the axial symmetry assumption for our Szekeres spacetimes \cite{C26}.

Finally, we will consider the possibility of higher multipoles (quadrupole, octupole, etc) in the Szekeres matter density. Currently, the few multipole measurements displayed in the literature give inconsistent results. While quadrupole and octupole components are claimed to have been measured in radio galaxy maps \cite{T19}, no evidence for power in higher multipoles than the quadrupole has been found in the CatWISE2020 quasar catalog \cite{vH26} and this quadrupole moment is warned to be consistent with noise by the authors. This suggests that new multipole searches are still needed to see whether the results suit or not the Szekeres expectations.

The paper is organized as follows. In Sec.\ref{sz}, the useful properties of the Szekeres solutions are recalled. The choice of the QSS solution as the sound class for reproducing the measured dipole is justified in Sec.\ref{qss}. A discussion enlightening the dipole issue in the Szekeres framework is displayed in Sec.\ref{dip}. The higher multipoles are studied in Sec.\ref{hm}. The tests using the mass angular features are described in Sec.\ref{multi}. Section \ref{co} is devoted to the conclusion.

\section{The Szekeres solution: a short reminder} \label{sz}

The solution found by Szekeres in 1975 \cite{S75} is an exact irrotational spacetime of General Relativity (GR), with no global symmetry, hence no Killing vector, and gravitationally sourced by dust, a zero-pressure perfect fluid. For a detailed presentation of the Szekeres solution the reader is referred to specialized textbooks \cite{BK10,P24} and references therein.

In synchronous and comoving projective coordinates, in geometric units, and using the parametrization proposed in \cite{H96}, the line element is
\begin{equation}
\textrm{d}s^2 = - \textrm{d}t^2 + \frac{\left(\Phi_{,r} - \Phi E_{,r}/E\right)^2}{\epsilon - k} \textrm{d}r^2 + \frac{\Phi^2}{E^2}(\textrm{d}p^2 + \textrm{d}q^2), \label{s1}
\end{equation}
with 
\begin{equation}
E(r,p,q) = \frac{S}{2}\left[ \left(\frac{p-P}{S}\right)^2 + \left(\frac{q-Q}{S}\right)^2 + \epsilon \right],  \label{s2}
\end{equation}
where $\Phi$ is a function of the $t$ and $r$ coordinates and $k$, $S$, $P$, and $Q$ are functions of $r$ only. The parameter $\epsilon$ determines whether the $(p,q)$ 2-surfaces of constant $\{t,r\}$ are 2-spheres $(\epsilon= +1)$, 2-pseudo-spheres, i. e., hyperboloids $(\epsilon = -1)$, or 2-planes $(\epsilon=0)$. 

The field equations with a cosmological constant reduce to:
\begin{equation}
\Phi_{,t}^2 = \frac{2 M}{\Phi} - k + \frac{\Lambda}{3} \Phi^2, \label{s3}
\end{equation}
\begin{equation}
4 \pi \rho(t,r,p,q) = \frac{M_{,r} - 3 M E_{,r}/E}{\Phi^2 (\Phi_{,r} - \Phi E_{,r}/E)}, \label{s4}
\end{equation}
where $M(r)$ is an integration function whose interpretation varies with the quasi-geometry of the considered region and $\rho$ is the matter energy density.

Equation (\ref{s3}) can be integrated as
\begin{equation}
t - t_B(r) = \int_0^{\Phi} \frac{\text{d}\tilde{\Phi}}{{\sqrt{\frac{2 M}{\tilde{\Phi}} - k + \frac{\Lambda}{3} \tilde{\Phi}^2}}}, \label{s5}
\end{equation}
where $t_B(r)$ is an arbitrary integration function, which stands for the Big Bang time in a cosmological context.

\section{Quasi-spherical versus quasi-hyperbolic Szekeres models} \label{qss}

In previous works, we have considered the possibility of representing the late universe, by a patchwork of any of the three classes of Szekeres models: QSS, QHS or quasi-plane (QPS) \cite{C24}. Henceforth, since a finite intrinsic dipole seems to have been measured in the matter distribution, we are led to cease considering quasi-plane and quasi-hyperbolic regions.

Indeed, in both classes, the 2-surfaces $\{t = \text{const}, r = \text{const}\}$ have an infinite surface which implies unbounded masses. Thus, the mass distribution is deprived of any finite dipole structure \cite{S75a,K12} which allows us to discard QHS and QPS models for our purpose.

We will therefore, in the following, consider only the QSS model.

\section{Dipolar decomposition of the densities} \label{dip}

As we have mentioned in Sec.\ref{intro}, two different monopole + dipole (or quasi-dipole) decomposition of the mass density in QSS can be found in the literature. The first has been displayed by Szekeres \cite{S75a} and the reasoning is perfectly proper. The second has been proposed by de Souza \cite{dS85} and widely used afterwards in a misleading way. Since the problem is rather subtle, we reproduce below both reasonings and correct or complete them where needed.

\subsection{Szekeres' analysis of the QSS surface density} \label{sd}

This analysis has been displayed by Szekeres in the months following the discovery of his solution. The corresponding paper \cite{S75a} has thus been written in the initial complex coordinate system. Here, for an easy comparison purpose, we give this reasoning in Hellaby's coordinates and parametrization \cite{H96}.

The surface density, $\sigma$, is defined such that the mass contained in the shell bounded by the spheres of radii $r$ and $r + \text{d}r$ is $\text{d}M = \text{d}r \iint \sigma \frac{\Phi^2}{E^2} \text{d}p\text{d}q$, which gives $\sigma = \rho \frac{\Phi_{,r} - \Phi E_{,r}/E}{\sqrt{1-k}}$. Inserting (\ref{s4}), we obtain
\begin{equation}
4 \pi \sigma = \frac{M_{,r} - 3M E_{,r}/E}{\Phi^2 \sqrt{1-k}}. \label{s6}
\end{equation}
Then, we make a coordinate change to polar coordinates defined by
\begin{equation}
(\frac{p - P}{S}, \frac{q - Q}{S}) = \cot \frac{\theta}{2} (\cos \phi, \sin \phi). \label{s7}
\end{equation} 
From (\ref{s2}), we obtain
\begin{equation}
\frac{E_{,r}}{E} = - \frac{S_{,r} \cos \theta + \sin \theta (P_{,r} \cos \phi + Q_{,r} \sin \phi)}{S}, \label{s8}
\end{equation}
which shows that $E_{,r}/E$ is a pure $\ell = 1$ function, i. e., a dipole vector $(Q_{,r}, P_{,r}, S_{,r})$ multiplied by the unit vector $(X = \sin \theta \sin \phi, Y = \sin \theta \cos \phi, Z = \cos \theta)$. Then, using (\ref{s8}) into (\ref{s6}), an expression for $\sigma$ follows as
\begin{equation}
\sigma = g(t,r)\left[1 + f(r) \cos \theta + g(r) \sin \theta \cos \phi + h(r) \sin \theta \sin \phi \right], \label{s9}
\end{equation}
with
\begin{eqnarray}
g(t,r) &=& \frac{M_{,r}}{4 \pi \Phi^2 \sqrt{1 - k}}, \quad f(r) = \frac{3 M S_{,r}}{S M_{,r}}, \nonumber \\ 
 \quad g(r) &=& \frac{3 M P_{,r}}{S M_{,r}}, \quad h(r) = \frac{3 M Q_{,r}}{S M_{,r}}. \label{s10}
\end{eqnarray}
Equation (\ref{s9}) can be written as
\begin{equation}
\sigma = \sigma_S + \Delta \sigma, \label{s11}
\end{equation}
with
\begin{equation}
\sigma_S = g(t,r), \label{s12a}
\end{equation}
\begin{equation}
\Delta \sigma = g(t,r)\left[f(r) \cos \theta + g(r) \sin \theta \cos \phi + h(r) \sin \theta \sin \phi \right]. \label{s12b}
\end{equation}
This shows that the surface density of a given $\{t = \text{const}, r = \text{const}\}$ shell can indeed be written as the sum of a monopole, $\sigma_S$ and a dipole, $\Delta \sigma$.

\subsection{De Souza's analysis of the mass density} \label{md}

A decade later, de Souza \cite{dS85} made an analogous analysis, but used the mass density, $\rho$, instead of the surface density. His result, after recasting in Hellaby's coordinates and parametrization, reads
\begin{equation}
\rho = \rho_S + \Delta \rho, \label{s13}
\end{equation}
with
\begin{equation}
4 \pi \rho_S = \frac{M_{,r} - 3 M \chi_{,r}/\chi}{\Phi^2(\Phi_{,r} - \Phi \chi_{,r}/\chi)}, \label{s14}
\end{equation}
\begin{equation}
4 \pi \Delta \rho = \frac{(\chi_{,r}/\chi - E_{,r}/E)(3 M \Phi_{,r} -  M_{,r}\Phi)}{\Phi^2 (\Phi_{,r} - \Phi E_{,r}/E)(\Phi_{,r} - \Phi \chi_{,r}/\chi)}, \label{s15}
\end{equation}
where
\begin{equation}
\chi = \frac{1 + P^2 + Q^2}{2 S} + \frac{S}{2}. \label{s16}
\end{equation}
In this decomposition, we see that $\rho_S$, being dependent only on $t$ and $r$ is constant on each $\{t = \text{const}, r = \text{const}\}$ shell and that it can be interpreted as a monopole. But the interpretation of $\Delta \rho$, as displayed by de Souza, has led to a number of misunderstanding in the sequel.

Restated by Peblanski and Krasi\'nski \cite{P24} in the Hellaby formalism, this interpretation can be formulated as follows. They show that the split given by (\ref{s13})-(\ref{s16}) implies that the surface $\Delta \rho = 0$ passes through the center of the shell. Then, they retain as a solution for $\Delta \rho = 0$, among the two available ones, that which has an equation independent of $t$. Thus, it is a world-sheet of a comoving surface, which they name $H_1$. It is obvious to see that $\Delta \rho$ changes sign when $H_1$ is crossed. The authors conclude therefore that $\Delta \rho$ is a ''dipole-like`` contribution to the matter density. Then, they verify that $H_1$ intersects every sphere of constant $t$ and $r$ along a circle parallel to the great circle $E_{,r} = 0$, unless the dipole component of the density vanishes. We can thus add to their reasoning that a circle passing through the center of the shell  and parallel to the great circle $E_{,r} = 0$ is itself the great circle $E_{,r} = 0$.

Anyhow, the fact that $\Delta \rho$ changes sign when a surface delimiting a great circle on the sphere is crossed is necessary but not sufficient for this quantity to be an exact dipole. We suspect this is the reason why de Souza, Peblanski, and Krasi\'nski call it a ''quasi-dipole``. However, in de Souza's paper, an unwelcome shortcut stating ''the mass distribution of the SQSS is of a monopole plus a dipole`` has mislead a number of following authors.

\subsection{Discussion} \label{dis}

Now, we will explain why $\Delta \rho$ is not a dipole, while $\Delta \sigma$ is. "Dipole" means that on each sphere $\{t = \text{const}, r = \text{const}\}$ the angular dependence is pure $\ell = 1$. That is a statement about the harmonic content, not the zero set. Any function with a zero-mean angular profile (quadrupole, octupole, a bump minus a dip, ...) also changes sign across some surface. Moreover, $\Delta \rho = 0$ having a solution that is a comoving world-sheet only tells one where the sign flips. It says nothing about the functional form of $\Delta \rho$ away from that sheet. A genuine dipole must also be odd under reflection through the "equator" and have zero average over the sphere.

In the formula (\ref{s4}) for $\rho$, as well as in (\ref{s15}) for $\Delta \rho$, $E_{,r}/E$ appears in both numerator and denominator, so $\rho$ and $\Delta \rho$ are Möbius-type functions of $E_{,r}/E$, i. e., ratios of linear functions of $E_{,r}/E$, e. g., for $\rho$,
\begin{equation}
\rho \propto \frac{M_{,r} - 3 M E_{,r}/E}{\Phi^2(\Phi_{,r} - \Phi E_{,r}/E)}. \label{s17}
\end{equation}

On a sphere, $E_{,r}/E$ contains only $\ell = 0$ and $\ell = 1$ harmonics, hence each term in the numerator and in the denominator above are also a sum of a $\ell = 0$ and of a $\ell = 1$ harmonics, but a ratio of two such combinations is not, so $\rho$, and therefore $\Delta \rho$, contain higher multipoles. Taken literally, ``$\Delta \rho$ is a dipole (or a quasi-dipole)'' is therefore only approximate or qualitative.

The mass, not the density, might justify this claim. The proper volume element is
\begin{equation}
\text{d} V \propto \frac{\Phi^2 (\Phi_{,r} - \Phi E_{,r}/E)}{E^2 \sqrt{1 - k}}\text{d}r \text{d}p \text{d}q. \label{s18}
\end{equation}
Thus the mass element reads
\begin{equation}
\text{d} m = \rho \text{d} V \propto \frac{M_{,r} - 3 M E_{,r}/E}{E^2} \text{d}r \text{d}p \text{d}q. \label{s19}
\end{equation}
Since $\text{d}p \text{d}q/E^2$ is proportional to the area element of the sphere, the mass per unit solid angle on each shell is linear in $E_{,r}/E$:
\begin{equation}
\frac{\text{d} m}{\text{d} \Omega} \propto M_{,r} - 3 M E_{,r}/E. \label{s20}
\end{equation}
Because, from (\ref{s8}), $E_{,r}/E$ is a pure dipole, the mass distribution over each sphere is exactly a monopole plus a dipole. It is a result about the mass distribution over each sphere, derived from this linearity, and not an inference from any sign change. The sign change across a comoving surface is a necessary, visible signature of a dipole. By itself it is not a proof. The rigorous statement is that the mass per solid angle on each $(t, r)$ sphere is exactly $\ell = 0$ plus $\ell = 1$. The same statement applies to the surface density, $\sigma$ for the same reasons, as shown by Szekeres \cite{S75a} and recalled in Sec.\ref{sd}. 

On the contrary, the quantity $\Delta \rho$, might only be named dipole-like, as coined by de Souza \cite{dS85} and picked up by Pleblanski and Krasi\'nski \cite{P24}. The extra dependence on $\Phi_{,r} - \Phi E_{,r}/E$ distorts it, and this distortion is also what produces shell crossings where the denominator vanishes, a feature with no counterpart in a pure dipole. In the dipole direction the mass is shifted to one side of the sphere. The sign of $M_{,r} - 3 M E_{,r}/E$, and hence whether the density is above or below the monopole value, flips across the world-sheet. The density $\rho$ is monopole plus quasi-dipole, with a sign-changing odd part. Dipole properties then hold only approximately. In particular, the deviation $\Delta \rho$ has an exactly odd numerator but not an exactly odd denominator. Its sphere average is not zero in general, and it contains multipoles at higher order.

\subsection{Axially symmetric case} \label{ax}

In a previous paper, we had proposed that the late universe might be axially symmetric according to some observations of an anisotropic expansion \cite{C26}.
We will therefore analyze what becomes of the above results in this particular case.

Axial symmetry means that the dipole direction is the same on every shell. As in \cite{C26}, we choose the angular coordinates such that $P = Q = 0$. Inserted in (\ref{s8}), this gives
\begin{equation}
\frac{E_{,r}}{E} = - \beta z, \label{s21}
\end{equation}
with  $\beta = S_{,r}/S$ and $z = \cos \theta$.
Therefore, the different quantities of interest become
\begin{equation}
\rho \propto \frac{M_{,r} + 3 M \beta z}{\Phi^2 (\Phi_{,r} + \Phi \beta z)}, \label{s22}
\end{equation}
\begin{equation}
\sigma \propto \frac{M_{,r} + 3M \beta z}{\Phi^2 \sqrt{1-k}}, \label{s23}
\end{equation}
\begin{equation}
\frac{\text{d} m}{\text{d} \Omega} \propto M_{,r} + 3 M \beta z. \label{s24}
\end{equation}
We see that, in the angular decomposition of $E_{,r}/E$, only $m = 0$ terms persist. There is no $\phi$ dependence, so $\rho$ expands in $P_{\ell}(\cos \theta)$ Legendre polynomials only. It is a Mobius-type function of $z$, i. e., a ratio of two linear functions in $z$. Its Legendre expansion contains every $\ell$. The higher multipoles come from the denominator $\Phi_{,r} + \Phi \beta z$, and their relative size is set by the dimensionless parameter $|\Phi \beta / \Phi_{,r}|$. At first order in that parameter, $\rho$ is monopole plus dipole. At second order a quadrupole appears, then, an octupole, and so on.

Indeed, in the axial case, with $j = \Phi \beta/ \Phi_{,r}$, the density has the form $\rho \propto (a + bz)/(1 + jz)$, where $a = M_{,r}$ and $b = 3 M \beta$. We can write it as follows:
\begin{equation}
\frac{a + bz}{1 + jz} = a + \frac{(b - a j) z}{1 + jz}. \label{s25}
\end{equation}
The monopole value is $\rho_S \propto a$ and the remainder is
\begin{equation}
\Delta \rho \propto \frac{z}{1 + jz} = z - j z^2 + j^2 z^3 - ..., \label{s26}
\end{equation}
which we split by parity under $z \rightarrow -z$. The odd part is
\begin{equation}
\frac{z}{1 - j^2z^2} = z + j^2 z^3 + ..., \label{s27}
\end{equation}
which contains only odd $\ell$s. The even part reads
\begin{equation}
\frac{-jz^2}{1 - j^2z^2} = - j z^2 + j^3 z^4 - .... \label{s28}
\end{equation}
This contains even $\ell$ only. 

We see that the corrections to $\rho=$ monopole + dipole can be small in the dimensionless parameter $j = \Phi S_{,r}/\Phi_{,r}S$. But this is not always the case. What is guaranteed is $|j| < 1$, since it is implied by the no-shell crossing conditions $E \Phi_{,r}/ \Phi - E_{,r} > 0$ for $\Phi_{,r} > 0$ or $E \Phi_{,r}/ \Phi - E_{,r} < 0$ for $\Phi_{,r} < 0$. 

The physical interpretation of $j$ is as follows. At any fixed $t$ coordinate, $\Phi_{,r} / \Phi = \partial \ln \Phi/\partial r$, so
\begin{equation}
j = \frac{\Phi S_{,r}}{S \Phi_{,r}} = \frac{\partial \ln S / \partial r}{\partial \ln \Phi/\partial r} = \frac{\text{d} \ln S}{\text{d} \ln \Phi}. \label{s29}
\end{equation}
It is dimensionless and measures how fast the dipole function $S$ changes from shell to shell, compared with how fast the areal radius $ \Phi$ changes. The denominator of $\rho$ is $\Phi_{,r} + \Phi \beta z = \Phi_{,r}(1 + jz)$, with $z = \cos \theta$ $\in$ [-1, 1]. It vanishes at $z = -1/j$, which lies on the sphere exactly when $|j| \geq 1$. Hence, the no-shell-crossing requirement implies $|j| < 1$. Therefore, the expansion $\Delta \rho \propto z/(1 + jz) = z - j z^2 + j^2z^3 - ...$ converges for $|j| < 1$. For $j$ close to 1 it converges slowly, and $\Delta \rho$ is strongly non-dipolar.

To proceed with our reasoning, we consider the Friedmann-like evolution equation (\ref{s3}). Near the Big-Bang, $\Phi \rightarrow 0$ and (\ref{s3}) becomes
\begin{equation}
\Phi_{,t}^2 = \frac{2M}{\Phi}, \label{s30}
\end{equation}
which can be written as
\begin{equation}
\sqrt{\Phi}\text{d}\Phi = \sqrt{2M}\text{d}t. \label{s31}
\end{equation}
Integrated from the Bang time, where $\Phi = 0$, this gives
\begin{equation}
\Phi = \left(\frac{9M}{2}\right)^{1/3}(t - t_B)^{2/3}. \label{s32}
\end{equation}
This is an Einstein-de Sitter-like growth, $\Phi \propto t^{2/3}$. At a fixed time $t$, the factor $(t - t_B)^{2/3}$ is constant if $t_B = \text{const}$. This is the choice that can be made to take into account the degree of freedom of the radial coordinate $r$ \cite{C24,C26}. In this case,
\begin{equation}
\Phi \propto M^{1/3}, \quad \text{so} \quad \frac{\Phi_{,r}}{\Phi} = \frac{M_{,r}}{3M}. \label{s32a}
\end{equation}
In these units and with this approximation -- negligible curvature -- $M$ is the mass inside the shells which expand as a flat dust universe with that mass. The mass and the radius are tied by $M \propto \Phi^3 (t - t_B)^2$, which implies that the mean density inside each shell is the same function of time for every shell. 

Now, since we have neglected above the curvature term, we must evaluate the magnitude of the corresponding uncertainty. The relative correction is $k \Phi/(2M)$. It scales as $k(t - t_B)^{2/3}/M^{2/3}$ and it is small only when the time $t$ is close to the Bang time, hence, in early cosmological times. For the radial derivative of $\Phi$, we have also to take $k_{,r}$ into account and this modifies $\Phi_{,r}/\Phi$ at the next order. Thus, early time means $t-t_{B} \ll M/|k|^{3/2}$.

We define $\zeta = (\Phi_{,r}/\Phi)/(M_{,r}/3M)$, the ratio of the radial growth of $\Phi$ to its early time value. Then, $j = (b/a)/\zeta$. At early times, for a simultaneous Bang $t_B = 0$, $\zeta=1$. Hence, $j = b/a$ and $\rho \propto (a + bz)/[1 + (b/a)z]$ which has no angular dependence. This agrees with the vanishing of the factor $3M \Phi_{,r} - M_{,r} \Phi$ on the numerator of $\Delta \rho$. In this model, the early universe is angularly uniform in density, while the surface density and the mass per solid angle exhibit a dipole. In this regime, $j$ equals the dipole over monopole ratio of the mass per solid angle, thus it is small if the dipole is small with respect to the monopole. For a positive mass on the whole sphere, with $M_{,r} > 0$, this implies $|b| < a$, which gives $|j| < 1$ at this epoch.

At later times, in an overdense region with growing structures, $\Phi_{,r}/\Phi$ becomes less than $M_{,r}/3 M$. So $\zeta < 1$ and $j$ grows. As a shell-crossing, where the model breaks down, approaches, $\Phi_{,r}/\Phi \rightarrow E_{,r}/E$ and $|j| \rightarrow 1$. Therefore, $j$ is small as soon as we are away from a shell-crossing.

Still in the late universe, when $\Phi_{,r}/\Phi > M_{,r}/3 M$, corresponding to a void case, $\zeta > 1$, and $j$ shrinks. 

The expansion in $j$ is a controlled approximation only when $|j| \ll 1$. This means a weak dipole ($|b/a| \ll 1$), at an epoch when $\zeta$ is far from 1, i. e., away from the Bang time. For a strong dipole, a “dipole-like” feature fails for $\rho$, though the mass per solid angle and the surface density stay exactly monopole plus dipole.

\section{Higher multipoles} \label{hm}

Now, we go back to the general QSS spacetime, not obligatorily axially symmetric, and consider the higher multipole moments of the mass density $\rho$. We denote $W = E_{,r}/E$. Let $u = (p-P)/S = \cot (\theta/2) \cos \phi$ and $v = (q-Q)/S = \cot (\theta/2) \sin \phi$. Inserting all this in (\ref{s8}), we obtain
\begin{equation}
W = A_0(r) + {\bf A}(r). \hat{\bf n}(\theta, \phi), \label{s33}
\end{equation}
with
\begin{equation}
A_0 = - \frac{S_{,r}}{S}, \quad {\bf A} = \left(-\frac{P_{,r}}{S}, -\frac{Q_{,r}}{S}, \frac{S_{,r}}{S} \right). \label{s34}
\end{equation}
$A_0$ is a monopole. ${\bf A}$ is a dipole vector whose direction defines a preferred direction on each sphere, ${\bf A}. \hat{\bf n} = |{\bf A}| \cos \gamma \equiv \mathcal{A} \cos \gamma P_1(\cos \gamma)$, with $P_{\ell}$s being Legendre functions of the first kind (here, $\ell = 1$).

Then, we define $N_0 \equiv M_{,r} - 3 M A_0$ and $D_0 \equiv \Phi_{,r} - \Phi A_0$. For a fixed $t$, $N_0$ and $D_0$ depend only on $r$. The whole angular dependence sits in $\mu \equiv \cos \gamma$ and the expression (\ref{s4}) of the density $\rho$ can be written as
\begin{equation}
4 \pi \rho = \frac{N_0 - 3 M \mathcal{A} \mu}{\Phi^2 (D_0 - \Phi \mathcal{A} \mu)}. \label{s35}
\end{equation}
Then, we set $\xi(r) \equiv \Phi  \mathcal{A}/D_0$. The inequality $|\xi| < 1$ corresponds to the no-shell-crossing condition $\Phi_{,r} - \Phi E_{,r}/E >0$. Now, we use the Legendre expansion of a simple pole:
\begin{equation}
\frac{1}{1 - \xi \mu} = \sum_{\ell=0}^{\infty} a_{\ell}\,P_\ell(\mu),\qquad |\xi|<1, \label{s36a}
\end{equation}
where
\begin{equation}
a_{\ell} = \frac{(2\ell+1)}{\xi}\,Q_\ell(\frac{1}{\xi}), \label{s36b}
\end{equation}
with $Q_{\ell}$ being Legendre functions of the second kind. Together with the Bonnet recursion formula:
\begin{equation}
(2\ell+1)\mu X_\ell = (\ell+1)X_{\ell+1}+\ell X_{\ell-1}, \label{s36c}
\end{equation}
where $X$ represents the Legendre functions of the first or second kind, this gives the exact multipole decomposition
\begin{equation}
\rho(r,\mu)=\sum_{\ell=0}^\infty \rho_\ell(r)\,P_\ell(\mu), \label{37}
\end{equation}
\begin{equation}
\rho_\ell(r)=\frac{(2\ell+1) Q_\ell(\frac{1}{\xi})(N_0 D_1 - N_1 D_0)}{4\pi \Phi^2 D_1^2}, \label{s38}
\end{equation}
where we have defined $N_1 \equiv 3 M \mathcal{A}$ and $D_1 \equiv D_0 \xi$.
Since $Q_\ell(x)$ has no real zeros for $x > 1$, $\rho_\ell(r) \neq 0$ for every $\ell$ provided $\mathcal{A}(r)\neq 0$ and $N_0 D_1 - N_1 D_0 \neq 0$. This last condition can be written equivalently as $M_{,r} \Phi - 3 M \Phi_{,r} \neq 0$. If this condition is not fulfilled and the equation becomes an equality, this would  imply that $M \propto \Phi^3$ locally. Moreover, $ M_{,r} \Phi - 3 M \Phi_{,r} = 0$, inserted in (\ref{s4}), gives $\rho = 3 M /(4 \pi \Phi^3 )$ which does not depend on $E_{,r}/E$. In this case, $\rho$ is the same at every point of the shell, for any dipole strength $\mathcal{A} \neq 0$, and the higher multipoles vanish. All the above is another proof that, provided the two conditions implying a non-vanishing numerator in $\rho_\ell$ are fulfilled, the density contains a full tower of multipoles, generated from the monopole $A_0$ and the dipole $\mathcal{A}(r)$.

The exact monopole is obtained by setting $\ell = 0$ in (\ref{s38}) that yields 
\begin{eqnarray}
\rho_0(r) &=& \frac{Q_0(1/\xi)}{4 \pi \Phi^2 D_1^2}(N_0 D_1 - N_1 D_0) \nonumber \\
&=& \frac{Q_0(1/\xi)}{4 \pi \Phi^3 \mathcal{A}} (M_{,r}\Phi - 3 M \Phi_{,r}), \label{s39}
\end{eqnarray}
with $Q_0 (1/\xi) = (1/2) \ln(\xi + 1)/(\xi - 1)$, which is completely different from the de Souza's result.

The actual dipole is obtained for $\ell = 1$ in (\ref{s38}), with $\ Q_1(1/\xi) = Q_0/\xi-1$, and reads
\begin{eqnarray}
&\rho_1(r)& = \frac{3\left(Q_0/\xi - 1\right)(N_0 D_1 - N_1 D_0)}{4 \pi \Phi^2 D_1^2} \nonumber \\
&=&\frac{3}{4 \pi \Phi^5 \mathcal{A}^3} \left[ Q_0 (\Phi_{,r} - \Phi A_o) - \Phi \mathcal{A}\right] (M_{,r} \Phi - 3 M \Phi_{,r}). \label{s39a}
\end{eqnarray}
We see that this expression for the dipole is fully different from that of the quasi-dipole given by (\ref{s15})-(\ref{s16}).

Now, we write the mass density as
\begin{equation}
\rho = \frac{1}{4 \pi \Phi^2}\left(\frac{N_1}{D_1} + \frac{N_0 D_1 - N_1 D_0}{D_1}\frac{1}{D_0 - D_1 \mu}\right), \label{s40}
\end{equation}
and we add the monopole and the dipole as given above. Naming $\rho_{ad} = \rho_0 + \rho_1$, we obtain
\begin{equation}
\rho_{ad} = \frac{1}{4 \pi \Phi^2 D_1^2} (N_0 D_1 - N_1 D_0) \left[\left(1 + \frac{3 D_0}{D_1} \right) Q_0 - 3\right]. \label{s41}
\end{equation}
It is easy to see that this sum is not equal to the mass density $\rho$, given by (\ref{s40}). Therefore, one cannot say that the mass density for a given shell is the sum of a monopole plus a dipole.

In the axisymmetric case, the global symmetry is achieved by setting $P_{,r} = Q_{,r} = 0$, and $\gamma= \theta$, the polar angle measured from the fixed symmetry axis. Every step of the derivation above is unchanged save for $\gamma \rightarrow \theta$ and $\xi = \Phi S_{,r}/(S D_0)$. 

Now, we will examine the behavior of the multipoles for $\ell > 1$, in the most general (non specifically axially symmetric) QSS case. From (\ref{s38}), we obtain
\begin{equation}
\frac{\rho_{\ell + 1}}{\rho_\ell}=\frac{(2 \ell + 3) Q_{\ell +1}\left(\frac{1}{\xi}\right)}{(2 \ell + 1) Q_{\ell}\left(\frac{1}{\xi}\right)}. \label{s42}
\end{equation}

Since $1/\xi > 1$, we can write $1/\xi = \cosh \eta$ and thus $\text{e}^{-\eta} = \xi/(1 + \sqrt{1 - \xi^2})$. 

Then, we divide the Bonnet recurrence formula (\ref{s36c}) by $\ell$ and let $\ell \rightarrow \infty$, to obtain the behavior of the $Q_{\ell}$ series for large $\ell$s. This gives
\begin{equation}
Q_{\ell + 1} - 2 x Q_{\ell} + Q_{\ell - 1} \simeq 0. \label{s43}
\end{equation}
Setting $x = 1/\xi$ and $\lambda_{\pm} = x \pm \sqrt{x^2 -1} = \text{e}^{\pm \eta}$, solution of the equation $\lambda^2 - 2 x \lambda + 1 = 0$, we recover (\ref{s43}) for $Q_{\ell} \propto \lambda^{\ell}$. The recurrence equation (\ref{s43}) has two independent solutions. The Legendre function of the first kind $P_{\ell}(x)$ which grows like $\text{e}^{\ell \eta}$ is the dominant solution. The Legendre function of the second kind $Q_{\ell} (x)$ which decays like $\text{e}^{-\ell \eta}$ is the minimal solution which is of interest for us \cite{P85}. By the Perron theorems \cite{P09,P10,P21}, the ratio of consecutive terms of the minimal solution tends to the smaller root:
\begin{equation}
\frac{Q_{\ell+1}}{Q_\ell} \rightarrow e^{-\eta},\quad \frac{2\ell+3}{2\ell+1} \rightarrow 1 \quad \Rightarrow\ \frac{\rho_{\ell+1}}{\rho_\ell} \rightarrow e^{-\eta}. \label{s44}
\end{equation}
The ratio test makes the distinction rigorous: $\rho_{\ell}$ is minimal if and only if $\rho_{\ell + 1}/ \rho_{\ell} \rightarrow \lambda_-$. This result obtains provided $|x| > 1$, which is the case with our definition of $x \equiv 1/\xi$ and $|\xi| <1$.

We can therefore conclude that the multipole amplitudes are exponentially decreasing functions of $\ell$ and that the first ones, such as the monopole and the dipole, are dominating over the following ones. Moreover, it appears that the smaller $\xi$ , the sharper the decrease. 

However, if $\xi \rightarrow 1^{-}$, i. e., the pole approaches $\mu =1$, $\gamma = 0$, and the decay rate vanishes. Hence, $\rho_{\ell}$ decreases only as $\sim \ln \ell$. If ever the shell-crossing is crossed, $|\xi| \geq 1$, $\rho$ diverges on the sphere and the Legendre expansion breaks down.

\section{The multipole tests} \label{multi}

A comoving observer located at $r_0$ on the symmetry axis of an axially symmetric QSS model surveys a $\{t = \text{const}, r_s = \text{const}\}$ 2-sphere. The cases $r_s = r_0$ and $r_0 = 0$ are special cases which will not be considered here. The two angles of interest are: $\theta$, the intrinsic polar angle on the sphere which labels the dust worldlines and $\psi$, the polar angle in the observer’s sky. Axial symmetry and an on-axis observer imply that rays have zero angular momentum about the axis, so the source azimuth equals the observed azimuth. In general $\psi \neq 0$ off the axis.

\subsection{The thin shell surface density} \label{ts}

We have seen in Sec.\ref{sd} that the shell surface density is exactly monopole plus dipole. From (\ref{s10}) and (\ref{s12b}), while considering that the dipoles in an axially symmetric QSS are aligned along the axis, i. e., $\theta = 0 \Rightarrow \cos \theta = 1$, we obtain that the dipole/monopole ratio is $s = 3MS_{,r}/ (SM_{,r})$.

A test using this surface density could be designed as follows. In a redshift shell extracted from a catalog, source masses per sky pixel are summed and divided by $D_A^2 \Omega_{\rm pix}$, where $\Omega_{\rm pix}$ is the solid angle of one pixel of the observer’s sky map. By the definition of the area distance, $D_A^2 \Omega_{\rm pix}$ is the cross-sectional area of the ray bundle that the pixel subtends, taken at the source and perpendicular to the line of sight. Here $D_A = D_A(z, \psi)$ is the direction-dependent area distance of that redshift shell. Then, expand this quantity in $P_{\ell}(\cos \psi)$.

Notice that the observed $\psi$-dipole is not the $\Delta \sigma$ above. Shells of constant $z$ are not shells of constant $r$, because $z$ depends on direction \cite{C24}. The sky coordinate $\psi$ also differs from $\theta$. Thus, the occurrence of $\ell \geq 2$ powers in this map does not contradict the ``exactly monopole + dipole'' proposition. It reflects the mapping $r \rightarrow z$. Therefore, it can be used to constrain the parameters of the Szekeres model. Comparing the sphere $\{t,z\}$ with the sphere $\{t,r\}$ requires the model, because the observer sees the past light cone and not a constant-t slice. The multipoles in the observed map must be transcribed in the Szekeres map with the null geodesic and the redshift equations \cite{C24} and the $\ell \geq 2$ powers in this map set to zero, hence the constraints on the Szekeres parameters.

For the sample to be axially symmetric and the comoving observer to be located on the axis, all the $a_{\ell, m}$ coefficients with $m \neq 0$ must vanish. Here, since the only non-zero multipole is the dipole, only the $a_{1,1}$ and $a_{1,-1}$ coefficients will have to be zero. This provides two other checks to be completed for a test of Szekeres axial symmetry with an on-axis observer.

\subsection{The mass per solid angle}

We consider $\Omega_{\rm obs}$, a solid angle on the observer’s celestial sphere, defined in the observer’s local rest frame such that $\text{d}\Omega_{\rm obs}$ is the standard area element of the unit 2-sphere in the 3-space orthogonal to the observer's 4-velocity. For a pixel map, $\Omega_{\rm pix}$ is the integral of $\text{d}\Omega_{\rm obs}$ over the pixel. The area distance is defined by $\text{d}A = D_A^2 \text{d} \Omega_{\rm obs}$. A test using this angle can be described as follows.

Measure $\text{d}m/\text{d}\Omega_{\rm obs}$ along each line of sight out to a given depth. For a dust model, the mass in the cone is exactly
\begin{equation}
\frac{\text{d}m}{\text{d}\Omega_{\rm obs}}=\int_{r_1(\psi)}^{r_2(\psi)}\!\text{d}r\;J(\psi,r)\,\frac{M_{,r}-3M\,E_{,r}/E}{4\pi\sqrt{1-k}}\Big|_{(r,\theta(\psi,r))}, \label{s45}
\end{equation}
with $\cos\Theta = (\varrho^2 - S^2)/(\varrho^2 + S^2)$ for the dust label at the event where the light ray crosses the shell, with $\varrho^2 = p^2 + q^2$ in the axially symmetric case. Three model-dependent maps enter:

- $r_i(\psi)$, from $z(r, \psi) = z_i$. A redshift slice cuts the shell at a direction-dependent $r$ coordinate.

- $\theta (\psi,r)$, the intrinsic polar angle where the ray crosses the shell $r$. The ray drifts across the dust worldlines, so each $r$ along it samples a different $(p, q)$.

- $J(\psi,r) = \text{d} \Omega _c / \text{d}\Omega_{\rm obs}$ at the crossing, where $\Omega_c$ is the solid angle on a shell’s own intrinsic unit sphere. On this sphere, the induced metric is $\Phi^2 (\text{d}p^2 + \text{d}q^2)/E^2$ and the area element is, thus, $1/E^2$. Hence,
\begin{equation}
\text{d} \Omega_c = \frac{\text{d}p \, \text{d}q}{E^2} = \sin \theta \,\text{d}\theta \, \text{d} \phi. \label{s46}
\end{equation}

All three maps come from the null geodesics of the model. The factor $\text{d}m/(\text{d}\Omega_c \text{d}r)$ itself is time-independent, so the crossing time $t(r, \psi)$ does not matter for it. Differentiating (\ref{s45}) with respect to $z$ gives the integrand $J (\text{d}r/\text{d}z)(M_{,r} - 3 M E_{,r}/E)/(4\pi\sqrt{1 - k})$, but the three maps remain. Without the model, only the cone integral is known. Thus to extract the shell dipole a QSS model must be assumed. A fit to the observations will therefore allow one to constrain the model parameters.

An on-axis observer simplifies the geometry a lot, but it doesn’t make the observed dipole equal to the dipole of the mass per solid angle. However, one can still complete the two different tests of axial symmetry described below. The dipole direction by redshift can be measured. This amounts to split the sample by redshift. Then a check that the dipole direction stays fixed can be performed. One can also use the property of the azimuthal modes by checking that the $a_{1,
\pm 1}$ coefficients are consistent with zero.

\subsection{The mass density}

Since we have shown that the smaller $\xi$, the sharper the decrease of the $\rho$ multipole amplitude, the measure of this decrease might be a measure of $\xi$ itself and could be used to set other constraints on the model parameters. 

A way of completing this is as follows. Measure the dipole of $\sigma$, which gives $s$, and the dipole of the radial proper thickness of the shell $(\Phi_{,r} + \Phi (S_{,r}/S)\cos \theta)\text{d}r/\sqrt{1-k}$, which gives $d = \Phi S_{,r}/(S \Phi_{,r})$. Then $\rho \propto (1+ s\cos \theta)/(1+ d\cos \theta)$, and the whole multipole tower follows from the formulas of Sec.\ref{hm} with $\xi = |d|$. In particular, (\ref{s44}) can be used to put constrains on the model.

\section{Conclusion} \label{co}

It is well-known that the Szekeres exact dust solution of General Relativity is properly designed to represent the late universe \cite{C24,C26}. Now, it has been claimed in the literature \cite{dS85,H02,B06,B09,BK10,K16,K18,K20,P24} that the mass energy density of the quasi-spherical class of the  Szekeres model can be uniquely and fully decomposed on a $\{t = \text{const}, r = \text{const}\}$ shell into a finite dipole living over a monopole. Thus, no other non-zero multipole components would be supposed to show up. We have shown here that this statement is misleading.

In the first part of this paper, we have corrected it, such as to obtain a set of exact propositions. We have considered three different densities: the surface density, $\sigma$, studied by Szekeres in 1975 \cite{S75a}, the mass density $\rho$, first analyzed by de Souza in 1985 \cite{dS85}, and the mass per solid angle considered here for the first time in this framework to our knowledge. We have shown that the surface density and the mass per solid angle on a shell $\{t = \text{const}, r = \text{const}\}$ can be indeed uniquely and fully decomposed into a monopole plus a dipole components and that no higher multipoles emerge there. On the contrary, the mass density $\rho$ is provided with a whole multipole tower, at variance with the standard statement.

After digging deeper into the angular properties of these different densities, and examining the particular features of the axially symmetric case, we have proposed a series of tests involving different cosmological catalogs able to yield number counts of adapted sources.

We have proposed, for each quantity, a test allowing to put constraints on the parameters of the underlying Szekeres model. While the thin shell surface density and the mass per solid angle both exhibit indeed a monopole + dipole only angular decomposition, this dipole is not the observed dipole. However, the observed dipole can be transcribed into the theoretical one by using the null geodesic and the redshift equations, and thus, the Szekeres parameters can be fitted.

As regards the mass density present in the stress-energy tensor, we have shown that the smaller some quantity $\xi$, the sharper the decrease of the $\rho$ multipole amplitudes. Hence, the measure of this decrease might be a measure of $\xi$ itself and could be used to set other constraints on the parameters of the model.

Following our proposal displayed and justified in \cite{C26} we have then considered an axially symmetric Szekeres model. An on-axis observer simplifies the geometry, but this does not make the observed dipole equal to the theoretical Szekeres dipole. Now, we have shown that, for the sample to be axially symmetric and the comoving observer to be located on the axis, all the $a_{\ell, m}$ coefficients with $m \neq 0$ must vanish. This has allowed us to design other checks to be completed for axial symmetry tests.

Since methods to measure the mass multipoles are currently being developed \cite{T19,vH26}, the upcoming releases of large galaxy surveys \cite{Ba20,H22,S26,C23,F26,J24} will allow us to complete the above proposed tests in the future. 

\bibliography{}

\end{document}